# STORAGE RING AND INTERACTION REGION MAGNETS FOR A $\mu^+\mu^-$ HIGGS FACTORY*

A.V. Zlobin#, Y.I. Alexahin, V.V. Kapin, V.V. Kashikhin, N.V. Mokhov, S.I. Striganov, I.S. Tropin, FNAL, Batavia, IL 60510, USA

*Abstract*

A low-energy Muon Collider (MC) offers unique opportunities to study the recently found Higgs boson. However, due to a relatively large beam emittance with moderate cooling in this machine, large-aperture high-field superconducting (SC) magnets are required. The magnets need also an adequate margin to operate at a large radiation load from the muon decay showers. General specifications of the SC dipoles and quadrupoles for the 125 GeV c.o.m. Higgs Factory with an average luminosity of ~2·10$^{31}$ cm$^{-2}$s$^{-1}$ are formulated. Magnet conceptual designs and parameters are reported. The impact of the magnet fringe fields on the beam dynamics as well as the IR and lattice magnet protection from radiation are also reported and discussed.

## INTRODUCTION

The major advantage of a $\mu+\mu-$ collider-based Higgs Factory (HF) [1] is the possibility of direct precise measurement of the Higgs boson mass and width. According to the general concept of such a machine [2], muons - in order to keep their longitudinal emittance and energy spread low - undergo only partial cooling leaving their transverse emittance relatively high (see Table 1). Therefore quite small values (a few cm) of the $\beta$-function at the Interaction Point (IP) are required resulting in a large beam size in the Final Focus (FF) quadrupoles.

Together with the necessity of magnets and detector protection from the muon decay showers, this leads to very large magnet apertures imposing challenging engineering constraints for magnet design and creating beam dynamics issues with magnet field quality and fringe fields.

Table 1: Higgs Factory Parameters

| Parameter | Unit | Value |
|---|---|---|
| Circumference | m | 300 |
| Beam energy | GeV | 62.5 |
| Transverse emittance, $\varepsilon_{\perp N}$ | ($\pi$)mm·rad | 0.3 |
| Longitudinal emittance, $\varepsilon_{\parallel N}$ | ($\pi$)mm·rad | 1.0 |
| Beam energy spread | % | 0.003 |
| $\beta^*$ | cm | 2.5 |
| Repetition rate | Hz | 30 |
| Muons/bunch | - | $2\times10^{12}$ |
| Average luminosity | cm$^{-2}$s$^{-1}$ | $2.5\times10^{31}$ |

Table 1 gives basic parameters of the HF Storage Ring (SR) preliminary design presented in [3]. The conceptual design and analysis of the Interaction Region (IR) large-aperture magnets has shown that the IR magnets with apertures ~500 mm and operation fields ~12 T are feasible using the Nb$_3$Sn technology.

___________________________________________
* Work supported by Fermi Research Alliance, LLC, under contract No. DE-AC02-07CH11359 with the U.S. Department of Energy and by the U.S. Department of Energy Muon Accelerator Program.
#zlobin@fnal.gov

This paper continues the design studies of the HF IR and SR magnets. The impact of the magnet field quality and fringe fields on the beam dynamics, the IR and lattice magnet protection from radiation are also discussed.

## STORAGE RING LATTICE

The HF lattice layout and the beam sizes are shown in Fig. 1 for parameters given in Table 1. The IR includes FF quadruplet (with first quadrupole Q1 at 3.5m from IP) and dipoles generating dispersion for subsequent 3-sextupole chromaticity correction [4]. The latter is necessary despite small energy spread to compensate the path length increase due to betatron oscillations. The matching section allows for $\beta^*$ variation in wide range.

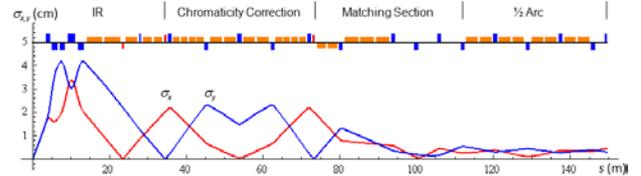

Figure 1: Layout and beam sizes in HF half-ring. Quadrupole magnets are shown in blue, dipoles in orange and sextupole correctors in red color.

## MAGNET REQUIREMENTS

Table 2 shows the IR magnet parameters discussed in [3]. The orbit sagitta in IR dipoles is quite large – 8.1 cm. However, it does not affect the IR dipole bore diameter since it is determined by the large vertical beam size.

Table 2: IR Magnet Specifications

| Parameter | Q1 | Q2 | Q3 | Q4 | B1 |
|---|---|---|---|---|---|
| 10$\sigma_{max}$ (mm) | 234 | 411 | 339 | 415 | 405 |
| G$_{nom}$ (T/m) | 74 | -36 | 44 | -25 | 0 |
| B$_{nom}$ (T) | 0 | 2 | 0 | 2 | 8 |
| L$_{mag}$ (m) | 1.00 | 1.40 | 2.05 | 1.70 | 4.10 |
| Coil aperture (mm) | 267 | 444 | 372 | 448 | 438 |
| Quantity | 1 | 2 | 1 | 1 | 2 |

Tables 3-4 present the maximum values of main parameters for the dipoles (B) and quadrupoles (Q) used in the chromaticity section (CS), the matching section (MS) and the arc (ARC). The most challenging magnets are the CS dipoles (B$_{CS}$), some MS dipoles (B$_{MS}$I) and the arc dipoles (B$_{ARC}$) which need high nominal operation field up to 10 T. This field level requires using the Nb$_3$Sn technology.

The magnet aperture outside the IR regions reduces from 231 mm in CS quadrupoles to 92 mm in arc dipoles. Note, that the aperture size in the arc is defined by the arc dipoles due to the relatively large beam sagitta. To standardize the magnet designs in different sections it was decided to use for this study two different aperture sizes – large ~231 mm in CS and adjacent part of MS, and small ~130 mm in ARC and adjacent part of MS.

Table 3: SR Dipole Magnet Specifications

| Parameter | $B_{CS}$ | $B_{MS}I$ | $B_{MS}II$ | $B_{ARC}$ |
|---|---|---|---|---|
| $10\sigma_{max}$ (mm) | 225 | 222 | 127 | 92 |
| $B_{nom}$ (T) | 10 | 10 | 6.4 | 10 |
| $L_{mag}$ (m) | 1.8 | 2.4 | 3.6 | 3.0 |
| Coil aperture (mm) | 258 | 255 | 160 | 120 |
| Quantity | 13 | 2 | 3 | 8 |

Table 4: SR Quadrupole Magnet Specifications

| Parameter | $Q_{CS}$ | $Q_{MS}$ | $Q_{ARC}$ |
|---|---|---|---|
| $10\sigma_{max}$ (mm) | 231 | 130 | 46 |
| $B_{nom}$ coil (T) | 5.3 | 5.5 | 3.3 |
| Length (m) | 1.0 | 1.0 | 1.0 |
| Coil aperture (mm) | 264 | 163 | 79 |
| Quantity | 5 | 5 | 4.5 |

## MAGNET DESIGNS AND PARAMETERS

The aperture of magnet coil in this study was defined as $10\sigma_{max}$ plus 30 mm for the beam pipe (BP) and absorber (ABS) plus 3 mm for BP insulation and helium channel. In the IR magnets the coil aperture was increased by ~50 mm with respect to this definition and limited by two sizes, 320 mm in Q1 and 500 mm in Q2-Q4 and B1, to reserve additional space for IR magnets protection against radiation [3]. The coil aperture of CS and adjacent MS magnets was increased to 270 mm, and in ARC magnets and adjacent MS magnets to 160 mm. This allows using thicker inner absorbers in arc magnets if necessary.

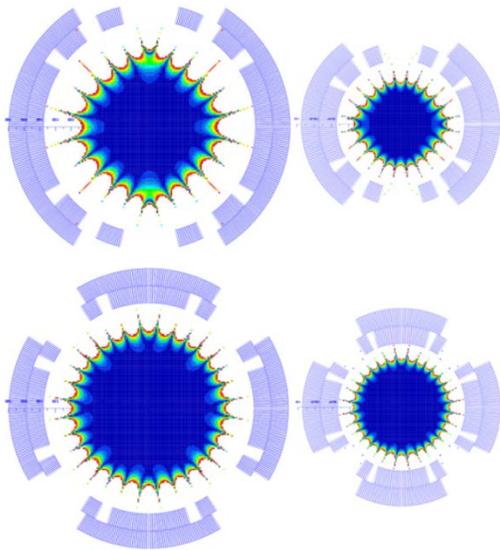

Figure 2: Coil cross-sections of HF SR dipoles (top) and quadrupoles (bottom) with 270 mm (left) and 160 mm (right) apertures.

The coil cross-sections were optimized with the ROXIE code [5] to achieve the required nominal operation field or gradient with appropriate margin and good field quality in the beam area of $8\sigma$. As in our previous studies [3] the conceptual designs of the HF SR magnets are based on a 42-strand Rutherford cable, 21.6 mm wide and 1.85 mm thick, made of a 1 mm strand with Cu/nonCu ratio of 1.0. The cable is insulated with a 0.2 mm thick insulation.

All magnets are based on 2-layer, shell-type coils with remote iron yoke used mainly to reduce fringe fields. The optimized D and Q coil cross-sections for CS, MS and ARC sections are shown in Fig. 2.

The main magnet parameters at 4.5 K are reported in Table 5. Relatively low field level in CS, MS and ARC quadrupoles allows using the traditional NbTi technology. Therefore, the parameters of these magnets were calculated for both conductor options, NbTi and $Nb_3Sn$.

Table 5: Nominal Magnet Parameters at $T_{op}$=4.5 K.

| Parameter | Dipole | | Quadrupole | |
|---|---|---|---|---|
| Coil ID (mm) | 270 | 160 | 270 | 160 |
| Max $B_{op}$ (T) | 10.0 | 10.0 | - | - |
| $B_{max}$ (T) | 14.1 | 14.2 | - | - |
| Max $G_{op}$ (T/m) | - | - | 33 | 36 |
| $G_{max}$ (T/m) | - | - | 54.1*/96.6# | 90.0*/160.8# |
| $B_{max}$ coil (T) | 15.8 | 15.6 | 8.7*/15.3# | 8.3*/14.7# |
| Fraction of SSL | 0.71 | 0.71 | 0.61*/0.34# | 0.4*/0.22# |
| L (mH/m) | 54.1 | 20.5 | 24.5 | 8.7 |
| $E_{op}$ (MJ/m) | 4.33 | 1.84 | 0.59 | 0.12 |
| $F_{x\_op}$ (MN/m) | 9.25 | 5.98 | 0.89 | 0.25 |
| $F_{y\_op}$ (MN/m) | -3.96 | -2.66 | -0.93 | -0.28 |

* - NbTi; # - $Nb_3Sn$

The $Nb_3Sn$ dipole magnets operate at 71% and the quadrupoles at 22-34% of their short sample limit (SSL) at 4.5 K. Using the same cable with NbTi strand in both quadrupoles reduces $G_{max}$ to 54 and 90 T/m. However, the magnets still have quite large operation margin, they operate at 61% and 40% of the SSL for 270 mm and 160 mm quads respectively. The final choice of the superconductor for SR quadrupoles will depend on the results of radiation studies for these magnets.

Geometrical field harmonics at the corresponding $R_{ref}$ are reported in Table 6. With optimized coil cross-sections the relative field errors in the area occupied by muon beams is $\sim 10^{-4}$ (dark blue area in Fig. 2).

Table 6: Geometrical Harmonics ($10^{-4}$).

| | ID, mm | $R_{ref}$, mm | $b_3$ | $b_5$ | $b_6$ | $b_7$ | $b_9$ | $b_{10}$ |
|---|---|---|---|---|---|---|---|---|
| **D** | 270 | 90 | -0.2 | -0.1 | - | 0.2 | 1.2 | - |
| | 160 | 53 | 0.0 | -0.1 | - | 0.5 | 1.1 | - |
| **Q** | 270 | 90 | - | - | 0.2 | - | - | 0.0 |
| | 160 | 53 | - | - | 0.1 | - | - | 0.0 |

The dynamic aperture (DA) on the plane of transverse particle positions at the IP and projection of the IR quad aperture onto this plane using linear optics functions is shown in Fig. 3. The DA analysis with MADX PTC code shows that field errors in the SR magnets reduce DA by a factor of 2 to a size of the magnet good field region.

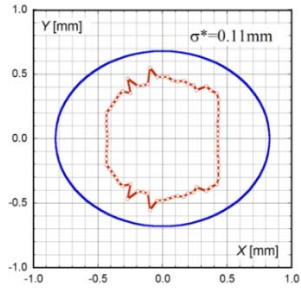

Figure 3: The dynamic aperture at IP and the projection of the IR quadrupole aperture (solid circle).

## ENERGY DEPOSITION IN MAGNETS

Energy deposition and detector backgrounds induced by decays of muons from both beams were simulated with MARS15 code [6] using our previous approach [7]. All the related details of geometry, materials and magnetic fields of the HF magnets in IR and SR were implemented in the 3D model along with a complete detector model, an optimized tungsten nozzle near IP and other components of the Machine-Detector Interface (Fig. 4).

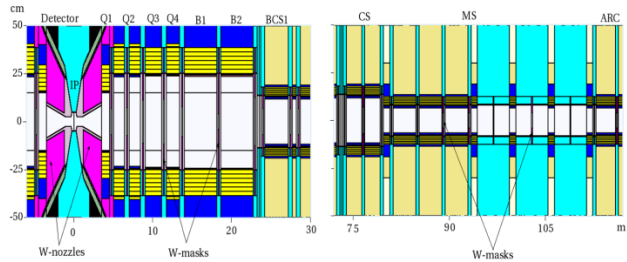

Figure 4: MARS model of HF MDI and SR.

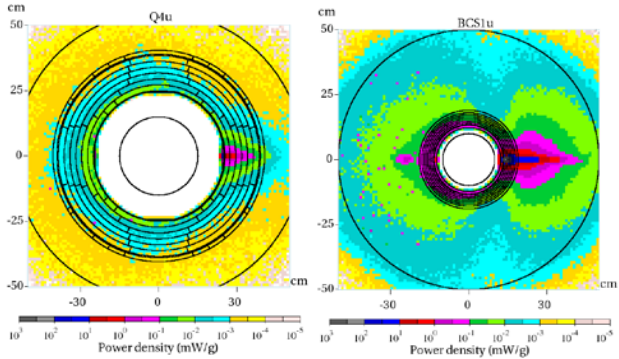

Figure 5: Power density (mW/g) in the IP ends of the Q4 quadrupole (left) and $B_{CS}1$ dipole (right).

To protect IR and SR magnets, 15 cm long, $10\sigma$ ID tungsten masks were installed in the magnet interconnect regions. Inner absorbers include up to 10-mm thick stainless steel liners inside each magnet with additional tungsten segments in both horizontal and vertical planes in some magnets along the ring.

Power density distributions were calculated for each of the HF collider ring magnet. The protection system performs adequately in the IR quadrupoles where the peak power density $P_p$ is at or below the quench limit of ~5 mW/g for $Nb_3Sn$ coils [3] (Fig. 5, left). However, it is ~10 times above the limit in the 500-mm coil ID B1 and B2 dipoles, and ~$10^2$ times above the limit for dipoles in CS, MS and ARC sections (e.g., Fig. 5, right). Additional tungsten inserts are needed to reduce $P_p$ below the limit.

Figure 6 shows the calculated dynamic heat load $W_d$ in IR and SR magnets for both muon beams. The presented $W_d$ values correspond to the entire magnet cross-sections including both cold and potentially warm elements. The typical $W_d$ value in SR magnets is ~1 kW/m as expected for the HF parameters of Table 1 except for the two dipoles at either end of the open section in MS and ARC, where it is ~2.5 kW/m. Note, that in spite of the low beam energy, the high number of decays in the HF keeps the level of $W_d$ high (~1 kW/m) as in the high energy machines [4], [7].

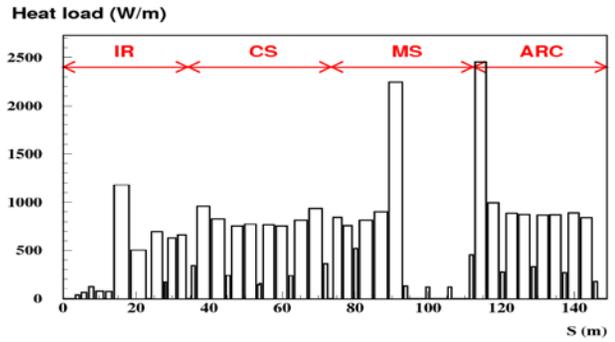

Figure 6: Dynamic heat load in IR and SR magnets.

## SUMMARY

Preliminary design and analysis of a Higgs Factory based on the 125 GeV c.o.m. $\mu^+\mu^-$ collider, including SR layout and beam dynamics, IR and SR magnet designs and parameters, have demonstrated the feasibility of such a machine. Protection of SR magnets from radiation and reduction of the dynamic heat load in cold part of magnet cryostat represent a serious problem and need further improvements.

## REFERENCES

[1] Y. Alexahin et al., "A Muon Collider Higgs Factory for Snowmass 2013", *arXiv:1308.2143v1 [hep-ph]* 9 Aug 2013; *Fermilab-Conf-13-245-T*.
[2] D.V. Neuffer et al., "A Muon Collider as a Higgs Factory", *Proc. of IPAC2013*, Shanghai, China, May 2013, p. 1472.
[3] A.V. Zlobin et al., "Preliminary Design of a Higgs Factory $\mu^+\mu^-$ Storage Ring", ibid, p. 1487.
[4] Y.I. Alexahin et al., "Muon Collider Interaction Region Design", *PRSTAB*, 14, 061001 (2011).
[5] ROXIE, http://cern.ch/roxie/
[6] MARS15, http://www-ap.fnal.gov/MARS/
[7] N.V. Mokhov et al., "Muon Collider Interaction Region and Machine-Detector Interface Design", *Proc. of PAC2011*, NYC, March 2011, p. 82.